\documentclass[11pt,nohyper,a4paper]{JHEP3}
\usepackage{amssymb,euscript,bbold}
\usepackage{amsmath,amscd}

\newcommand{\sect}[1]{\setcounter{equation}{0}\section{#1}}

\newcommand{\be}{\begin{eqnarray}}
\newcommand{\ee}{\end{eqnarray}}
\newcommand{\bea}{\begin{eqnarray}}
\newcommand{\eea}{\end{eqnarray}}
\newcommand{\ba}{\begin{array}}
\newcommand{\ea}{\end{array}}
\newcommand{\nn}{\nonumber \\}

\newcommand{\bR}{\mathbb{R}}

\title{General Supersymmetric Solutions of Five-Dimensional Supergravity}

\author{Jan B. Gutowski \\ DAMTP, Centre for Mathematical Sciences\\
University of Cambridge\\
Wilberforce Road, Cambridge, CB3 0WA, UK}

\author{Wafic Sabra \\ Centre for Advanced Mathematical Sciences and Physics Department\\
American University of Beirut\\ Lebanon}

\abstract{The classification of 1/4-supersymmetric solutions of five dimensional
gauged supergravity coupled to arbitrary many abelian vector multiplets,
which was initiated in \cite{gutowskireall:04}, is completed. The
structure of all solutions for which the Killing vector
constructed from the Killing spinor is null is investigated in
both the gauged and the ungauged theories and some new solutions
are constructed.}

\keywords{Supergravity Models}

\preprint{}

\begin{document}

\setcounter{equation}{0}

\sect{Introduction}

Recently there has been a lot of interest in the study of black holes in
gauged supergravity theory. This has been to a large extent motivated by the
conjectured equivalence between string theory on anti-de~Sitter (AdS) spaces
and certain superconformal gauge theories living on the boundary of AdS
\cite{maldacena:98, witten:98, oz:98}. From the point of view of the dual
CFT, supergravity vacua could correspond to an expansion around non-zero
vacuum expectation values of certain operators, or describe a holographic
renormalization group flow \cite{deBoer:1999xf}. It is hoped that such
equivalence will allow some understanding of the nonperturbative structure
of these gauge theories by studying classical supergravity solutions. An
example in this direction is the Hawking-Page phase  transition \cite{hp}
which is interpreted as a thermal phase transition from a confining to a
deconfining phase in the dual four dimensional $N=4$ supersymmetric
Yang-Mills theory \cite{wit}.

In this paper we will concentrate on five-dimensional $N=2$ gauged
supergravity coupled to vector multiplets and the classification of their
supersymmetric solutions. These are relevant for the holographic
descriptions of four dimensional field theories with less than maximal
supersymmetry. In the past few years, a lot of effort has been devoted to
finding solutions of these theories. For example, magnetically charged
string solutions preserving a quarter of supersymmetry were given in
\cite{cs1, ks1}. Also, supersymmetric electric solutions preserving half of
supersymmetry have been discussed in \cite{behrndt:98, klemm:01a} . However,
those electric and rotating solutions have naked singularities or naked
closed timelike curves. Some black string solutions and domain wall
solutions with non-trivial scalar fields were also presented in~\cite{cks1}.

More recently a systematic approach has been used to classify supersymmetric
solutions of the minimal gauged five
dimensional supergravity \cite{gauntlett:03}.
This approach was first used by Tod \cite{tod:83} for the
classification of supersymmetric solutions of minimal $\mathcal{N}=2$, $D=4$
supergravity. The basic idea in this analysis is to assume the existence of
a Killing spinor, (i.e., to assume that the solution preserves at least one
supersymmetry) and construct differential forms as bilinears in the Killing
spinor. These forms satisfy algebraic and differential conditions, which are
sufficient to determine the local form of the metric and the bosonic fields
in the theory. In \cite{gauntlett:03}, the solutions fall into two classes
depending on whether the Killing vector constructed from the Killing spinor
is null or time-like. It must be stressed that this general framework can
generate many new interesting solutions which are not easily found by
employing the usual method of simply guessing an Ans\"{a}tze. Among
other explicit solutions, supersymmetric asymptotically anti-de Sitter black
hole solutions were constructed for the minimal supergravity theory in
\cite{gutowski:04} and later generalised to the $U(1)^{3}$
theory (with three $R$ -charges) in \cite{gutowskireall:04}.
Further generalizations of these black holes have also recently been
found in \cite{cveticpope1:05} and \cite{cveticpope2:05}.
These solutions must have non-vanishing
angular momentum and unlike the solutions of \cite{klemm:01a} do not have
naked closed timelike curves. Moreover, in \cite{gutowskireall:04}, explicit
algebraic and differential equations were derived for the case where the
Killing vector is time-like and the
scalar fields take values in symmetric spaces \cite{gunaydin:85}.

In this work, the classification which was initiated in
\cite{gutowskireall:04} is completed. In particular, we relax the
requirement that the scalar manifold should be symmetric. The
constraint equations in the case for which the
Killing vector is time-like are derived.
The structure of all solutions with null Killing vector is also
investigated in both gauged and ungauged supergravity theories in five
dimensions with some explicit solutions given. We organise our work as
follows. Section two includes a brief review of the theories of $N=2$,
$D=5$ gauged supergravity coupled to $n$ abelian vector multiplets, the
equations of motion and the general analysis of the algebraic and
differential properties of the differential forms constructed from a
commuting Killing spinor \cite{gutowskireall:04}. In section three we will
analyse the case where the Killing vector is time-like and where the scalars
are not necessarily living in a symmetric space. Section four contains the
analysis of the solutions with null Killing vector in both gauged and
ungauged theories. Solutions are constructed which contain the
solutions of \cite{cks1} as a subclass. We present our
conclusions in section five.

\sect{Supersymmetric solutions of $N=2$ supergravity}

\label{section2}

\subsection{$N = 2$ supergravity}

\label{section2.1}

The action of $N=2$, $D=5$ gauged supergravity coupled to $n$
abelian vector multiplets is~\cite{gunaydin:85}
\begin{eqnarray}
S = {\frac{1 }{16 \pi G}} \int \big( -{}^5 R + 2 \chi^2 {\mathcal{V}}
-Q_{IJ} F^I \wedge *F^J +Q_{IJ} dX^I \wedge \star dX^J  \notag \\
-{\frac{1 }{6}} C_{IJK} F^I \wedge F^J \wedge A^K \big)
\end{eqnarray}
where $I,J,K$ take values $1, \ldots ,n$ and $F^I=dA^I$.
The metric has mostly negative signature. $C_{IJK}$ are
constants that are symmetric on $IJK$; in this paper we shall not assume
that the $C_{IJK}$ satisfy the non-linear ``adjoint identity''
which arises when the scalars lie in a symmetric space; though
we will assume that $Q_{IJ}$ is invertible, with inverse $Q^{IJ}$.

The $X^I$ are scalars which are constrained via
\begin{equation}  \label{eqn:conda}
{\frac{1 }{6}}C_{IJK} X^I X^J X^K=1\,.
\end{equation}
We may regard the $X^I$ as being functions of $n-1$ unconstrained scalars
$\phi^a$. It is convenient to define
\begin{equation}
X_I \equiv {\frac{1 }{6}}C_{IJK} X^J X^K
\end{equation}
so that the condition~({\ref{eqn:conda}}) becomes
\begin{equation}
X_I X^I =1\,.
\end{equation}
In addition, the coupling $Q_{IJ}$ depends on the scalars via
\begin{equation}
Q_{IJ} = {\frac{9 }{2}} X_I X_J -{\frac{1 }{2}}C_{IJK} X^K
\end{equation}
so in particular
\begin{equation}
Q_{IJ} X^J = {\frac{3 }{2}} X_I\,, \qquad Q_{IJ} \partial_a X^J = -{\frac{3
}{2}} \partial_a X_I\,.
\end{equation}
The scalar potential can be written as
\begin{equation}
{\mathcal{V}} = 9 V_I V_J (X^I X^J - {\frac{1 }{2}} Q^{IJ})
\end{equation}
where $V_I$ are constants.

For a bosonic background to be supersymmetric there must be a spinor
\footnote{ We use symplectic Majorana spinors. Our conventions are the same
as \cite {gauntlett:02}.}
$\epsilon^a$ for which the supersymmetry variations of the
gravitino and dilatino vanish. For the gravitino this requires
\begin{equation}  \label{eqn:grav}
\left[\nabla_\mu +{\frac{1 }{8}}X_I(\gamma_\mu{}^{\nu \rho} -4 \delta_\mu{}
^\nu \gamma^\rho) F^I{}_{\nu \rho} \right] \epsilon^a-{\frac{\chi }{2}} V_I
(X^I \gamma_\mu-3 A^I{}_\mu) \epsilon^{ab} \epsilon^b=0
\end{equation}
and for the dilatino it requires
\begin{equation}  \label{eqn:dil}
\left[ \left( {\frac{1 }{4}}Q_{IJ} \gamma^{\mu \nu} F^J{}_{\mu \nu}
+{\frac{3 }{4}} \gamma^\mu \nabla_\mu X_I \right) \epsilon^a
-{\frac{3 \chi }{2}}
V_I \epsilon^{ab} \epsilon^b \right] {\frac{\partial X^I }{\partial \phi^a}}
= 0\,.
\end{equation}
The Einstein equation is
\begin{equation}  \label{eqn:ein}
{}^5 R_{\alpha \beta} +Q_{IJ} F^I{}_{\alpha \lambda} F^J{}_\beta{}^\lambda
-Q_{IJ} \nabla_\alpha X^I \nabla_\beta X^J-{\frac{1 }{6}}g_{\alpha \beta}
\left(4 \chi^2 V +Q_{IJ} F^I{}_{\mu \nu} F^{J \mu \nu} \right) =0\,.
\end{equation}
The Maxwell equations (varying $A^I$) are
\begin{equation}  \label{eqn:gauge}
d \left(Q_{IJ} \star F^J \right)=-{\frac{1 }{4}}C_{IJK} F^J \wedge F^K\,.
\end{equation}
The scalar equations (varying $\phi^a$) are
\begin{eqnarray}  \label{eqn:scal}
\bigg[{-}d ( \star dX_I )+ \bigg( X_M X^P C_{NPI}-{\frac{1 }{6}} C_{MNI}
\bigg) (F^M \wedge \star F^N - dX^M \wedge \star dX^N )  \notag \\
- {\frac{3 }{2}} \chi^2 V_M V_N Q^{ML} Q^{NP} C_{LPI} \mathrm{dvol} \bigg]
{\frac{\partial X^I }{\partial \phi^a}} = 0\,.\qquad
\end{eqnarray}
If a quantity $L_I$ satisfies $L_I \partial_a X^I = 0$ then there must be a
function $M$ such that $L_I = M X_I$. This implies that the dilatino
equation~({\ref{eqn:dil}}) can be simplified to
\begin{equation}  \label{eqn:newdil}
\left[ \left({\frac{1 }{4}}Q_{IJ}-{\frac{3 }{8}}X_I X_J \right) F^J{}_{\mu
\nu} \gamma^{\mu \nu} +{\frac{3 }{4}} \gamma^\mu \nabla_\mu X_I \right]
\epsilon^a +{\frac{3 \chi }{2}}\left(X_I V_J X^J -V_I \right) \epsilon^{ab}
\epsilon^b =0\,,
\end{equation}
and the scalar equation can be written as
\bea
&-&d \left(\star dX_I \right) + \bigg({\frac{1 }{6}} C_{MNI} -{\frac{1 }{2}}
X_I C_{MNJ} X^J \bigg) dX^M \wedge \star dX^N   \nn
+ \bigg( X_M X^P C_{NPI}&-&{\frac{1 }{6}}C_{MNI}-6 X_I X_M X_N+{\frac{1 }{6}}
X_I C_{MNJ} X^J \bigg) F^M \wedge \star F^N   \nn
&+& 3 \chi^2 \bigg( {\frac{1 }{2}} V_M V_N Q^{ML} Q^{NP} C_{LPI} + X_I Q^{MN}
V_M V_N
\nn
&-&2 X_I X^M X^N V_M V_N \bigg) \mathrm{dvol} =0 \ .
\eea

\subsection{General supersymmetric solutions}

\label{section2.3}

Following~\cite{gauntlett:02}, our strategy for determining the general
nature of bosonic supersymmetric solutions is to analyse the differential
forms that can be constructed from a (commuting) Killing spinor. We first
investigate algebraic properties of these forms, and then their differential
properties.

From a single commuting spinor $\epsilon^a$ we can construct a scalar $f$, a
1-form $V$ and three 2-forms $\Phi^{ab} \equiv \Phi^{(ab)}$:
\begin{equation}
f \epsilon^{ab} = \bar{\epsilon}^a \epsilon^b\,, \qquad V_\alpha
\epsilon^{ab} = \bar{\epsilon}^a \gamma_\alpha \epsilon^b\,, \qquad
\Phi^{ab}_{\alpha \beta} = \bar{\epsilon}^a \gamma_{\alpha \beta}
\epsilon^b\,.
\end{equation}
$f$ and $V$ are real, but $\Phi^{11}$ and $\Phi^{22}$ are complex conjugate
and $\Phi^{12}$ is imaginary. It is convenient to work with three real
two-forms $J^{(i)}$ defined by
\begin{equation}
\Phi^{(11)} = J^{(1)} + i J^{(2)}\,, \qquad \Phi^{(22)} = J^{(1)} - i
J^{(2)}\,, \qquad \Phi^{(12)} = - i J^{(3)}\,.
\end{equation}
It will be useful to record some of the algebraic identities that can be
obtained from the {}Fierz identity:
\begin{eqnarray}  \label{eqn:XcontX}
V_{\alpha} V^{\alpha} &=& f^2\,, \nn
J^{(i)} \wedge J^{(j)} &=& -2\delta_{ij} f \star V\,, \nn
i_V J^{(i)} &=& 0\,, \nn
i_V \star J^{(i)} &=& - f J^{(i)}\,, \nn
J^{(i)}_{\gamma \alpha} J^{(j) \gamma}{}_{\beta} &=& \delta_{ij} \left( f^2
\eta_{\alpha\beta} - V_{\alpha} V_{\beta} \right) + \epsilon_{ijk} f
J^{(k)}_{\alpha\beta}
\end{eqnarray}
where $\epsilon_{123} = +1$ and, for a vector $Y$ and $p$-form $A$, $(i_Y
A)_{\alpha_1, \ldots, \alpha_{p-1}} \equiv Y^{\beta} A_{\beta \alpha_1,
\ldots, \alpha_{p-1}}$. {}Finally, we have
\begin{eqnarray}  \label{eqn:Phiproja}
V_{\alpha} \gamma^\alpha \epsilon^a &=& f \epsilon^a\
\end{eqnarray}
and
\begin{eqnarray} \label{eqn:Phiprojb}
\Phi^{ab}_{\alpha \beta} \gamma^{\alpha \beta} \epsilon^c &=& 8f
\epsilon^{c(a} \epsilon^{b)} \,.
\end{eqnarray}
Equation~(\ref{eqn:XcontX}) implies that $V$ is timelike, null or zero. The
final possibility can be eliminated using arguments in
\cite{gauntlett:02,reall:03}.

We now turn to the differential conditions that arise because $\epsilon$ is
a Killing spinor. We differentiate $f$, $V$, $\Phi$ in turn and use
({\ref{eqn:grav}}). Starting with $f$ we find
\begin{equation}  \label{eqn:df}
df = -i_V \left(X_I F^I \right),
\end{equation}
which implies $\mathcal{L}_V f=0$ where $\mathcal{L}$ denotes the Lie
derivative. Next, differentiating $V$ gives
\begin{equation}
D_{(\alpha} V_{\beta)} = 0\,,
\end{equation}
so $V$ is a Killing vector, and
\begin{equation}  \label{eqn:dV}
dV = 2f X_I F^I + X_I \star (F^I \wedge V) +2 \chi V_I X^I J^{(1)}\,.
\end{equation}
Finally, differentiating $J^{(i)}$ gives
\begin{eqnarray}  \label{eqn:dPhi}
D_\alpha J^{(i)}_{\beta\gamma} &=& -{\frac{1 }{2}} X_I \left[ 2
F^I{}_{\alpha}{}^{\delta} \left( \star J^{(i)} \right)_{\delta\beta\gamma}
-2 F^I{}_{[\beta}{}^{\delta} \left( \star J^{(i)} \right)_{\gamma] \alpha
\delta} + \eta_{\alpha [\beta} {}F^{I \delta \epsilon} \left( \star J^{(i)}
\right)_{\gamma] \delta\epsilon} \right]  \notag \\
&&- 2 \chi V_I X^I \delta^{i1} \eta_{\alpha [\beta} V_{\gamma]} +3 \chi
\epsilon^{1ij}V_I \left[ A^I{}_\alpha J^{(j)}{}_{\beta \gamma} +{\frac{1 }{3}
} X^I (\star J^{(j)}){}_{\alpha \beta \gamma} \right],
\end{eqnarray}
which implies
\begin{equation}  \label{eqn:cclos}
dJ^{(i)} = 3 \chi \epsilon^{1ij} V_I \left( A^I \wedge J^{(j)}+X^I \star
J^{(j)} \right)
\end{equation}
so $dJ^{(1)}=0$ but $J^{(2)}$ and $J^{(3)}$ are only closed in the ungauged
theory (i.e.\ when $\chi=0$). Equation~(\ref{eqn:cclos}) implies
\begin{equation}  \label{eqn:lied}
\mathcal{L}_V J^{(i)} = 3 \chi \epsilon^{1ij}\left(i_V (V_I A^I) -V_I X^I f
\right) J^{(j)} \,.
\end{equation}
Now consider the effect of a gauge transformation $A^I \rightarrow
A^I+d\Lambda^I$. The Killing spinor equation is invariant provided the
spinor transforms according to
\begin{eqnarray}  \label{eqn:kspmap}
\epsilon^1 &\rightarrow& \cos \left({\frac{3 \chi V_I \Lambda^I }{2}}
\right) \epsilon^1 - \sin \left({\frac{3 \chi V_I \Lambda^I }{2}} \right)
\epsilon^2  \notag \\
\epsilon^2 &\rightarrow& \cos \left({\frac{3 \chi V_I \Lambda^I }{2}}
\right) \epsilon^2 + \sin \left({\frac{3 \chi V_I \Lambda^I }{2}} \right)
\epsilon^1 \,.
\end{eqnarray}
Under these transformations, $f \rightarrow f$, $V \rightarrow V$ and
$J^{(1)} \rightarrow J^{(1)}$, but $J^{(2)} +i J^{(3)} \rightarrow e^{-3i
\chi V_I \Lambda^I} (J^{(2)}+i J^{(3)})$, so $J^{(2,3)}$ are only
gauge-invariant in the ungauged theory. We shall choose to work in a gauge
in which
\begin{equation}  \label{eqn:gaugechoice}
i_V A^I =f X^I\,.
\end{equation}
In such a gauge we have $\mathcal{L}_V J^{(i)}=0$.

To make further progress we will examine the dilatino equation
({\ref{eqn:newdil}}). Contracting with ${\bar{\epsilon}}^c$ we obtain
\begin{equation}  \label{eqn:liescal}
\mathcal{L}_V X_I =0
\end{equation}
and
\begin{equation}  \label{eqn:mostuseful}
\left({\frac{1 }{4}} Q_{IJ} -{\frac{3 }{8}} X_I X_J \right) F^J{}_{\mu \nu}
(J^{(i)} )^{\mu \nu} =- {\frac{3 \chi }{2}} \delta^{1i} (X_I V_J X^J -V_I )
f\,.
\end{equation}
Next, contracting~({\ref{eqn:newdil}}) with ${\bar{\epsilon}}^c
\gamma^\sigma $ we find
\begin{equation}  \label{eqn:liegaugeb}
i_V F^J = -d (f X^J )\,,
\end{equation}
which implies that
\begin{equation}
\mathcal{L}_V F^J=0\,.
\end{equation}
Hence $V$ generates a symmetry of all of the fields. In the gauge
({\ref{eqn:gaugechoice}}) we also have
\begin{equation}
\mathcal{L}_V A^I =0\,.
\end{equation}
Contracting~({\ref{eqn:newdil}}) with ${\bar{\epsilon}}^c \gamma^\sigma$ we
obtain the identity
\begin{equation}  \label{eqn:mostusefulb}
-\bigg({\frac{1 }{4}} Q_{IJ} -{\frac{3 }{8}} X_I X_J \bigg) F^J{}_{\mu \nu}
( \star J^{(i)} )_\sigma{}^{\mu \nu} =-{\frac{3 }{4}} (J^{(i)})_\sigma{}^\mu
\nabla_\mu X_I -{\frac{3 \chi }{2}} \delta^{i1} (X_I V_J X^J -V_I ) V_\sigma
\,.
\end{equation}
Finally, contracting~({\ref{eqn:newdil}}) with ${\bar{\epsilon}}^c
\gamma^{\sigma \lambda}$ gives {\footnote{These equations correct
some typographical errors found in equations (2.53) and (2.54) of
\cite{gutowskireall:04}, though these equations were not actually used
in the analysis of that paper.}}
\begin{eqnarray}  \label{eqn:mostusefulc}
\left( {\frac{1 }{4}} Q_{IJ} -{\frac{3 }{8}} X_I X_J \right) (F^J{}_{\mu
\nu} (\star V)^{\sigma \lambda \mu \nu}+2 f F^{J \lambda \sigma} ) =-{\frac{
3 }{4}}(\nabla^\lambda X_I V^\sigma - \nabla^\sigma X_I V^\lambda )  \notag
\\
+ {\frac{3 \chi }{2}} (X_I V_J X^J -V_I ) (J^{(1)})^{\sigma \lambda}\qquad
\end{eqnarray}
and
\begin{eqnarray}  \label{eqn:mostusefuld}
\left({\frac{1 }{2}} Q_{IJ} -{\frac{3 }{4}} X_I X_J \right) \left(F^{J
\sigma}{}_\nu (J^{(i)})^{\nu \lambda} - F^{J \lambda}{}_\nu (J^{(i)})^{\nu
\sigma}\right) = {\frac{3 }{4}} \nabla_\mu X_I (\star J^{(i)})^{\sigma
\lambda \mu}  \notag \\
+ {\frac{3 \chi }{2}} \epsilon^{1ij} \big(X_I V_J X^J -V_I \big)
\left(J^{(j)}\right)^{\sigma \lambda}\,.
\end{eqnarray}

\sect{The timelike case}

As in~\cite{gauntlett:02, gauntlett:03}, it is useful to consider two
classes of solution, depending on whether the scalar $f$ vanishes
everywhere or not. In the
null class, the vector $V$ is globally a null Killing vector. In the
timelike class,  there is some open set $\mathcal{U}$ in which $f$ is
non-vanishing and hence $V$ is a timelike Killing vector field,
and without loss of
generality one can take $f>0$ in $\mathcal{U}$~\cite{gauntlett:02}.
We first analyse the timelike class be examining the constraints
imposed by supersymmetry in the region $\mathcal{U}$. This analysis is very
similar to that presented in \cite{gutowskireall:04} for the case in which
the scalars lie in a symmetric space, and the $C_{IJK}$ satisfy an
additional non-linear algebraic constraint. We will not assume this here
however, unless stated explicitly.

Introduce coordinates $(t,x^m)$ such that $V = \partial/\partial t$. The
metric can then be written locally as
\begin{equation}  \label{eqn:metric}
ds^2=f^2(dt+\omega)^2-f^{-1}h_{mn}dx^m dx^n\,.
\end{equation}
The metric $h_{mn}$ can be regarded as the metric on a four dimensional
Riemannian manifold, which we shall refer to as the ``base space" $B$.
$\omega$ is a 1-form on $B$. Since $V$ is Killing, $f$, $\omega$ and $h$ are
independent of $t$. We shall reduce the necessary and sufficient conditions
for supersymmetry to a set of equations on $B$. Let
\begin{equation}  \label{eqn:e0def}
e^0 = f (dt+\omega)\,.
\end{equation}
We choose the orientation of $B$ so that $e^0 \wedge \eta_4$ is positively
oriented in five dimensions, where $\eta_4$ is the volume form of $B$. The
two form $d\omega$ can be split into self-dual and anti-self-dual parts on
$B$:
\begin{equation}  \label{eqn:rsp}
f d\omega=G^{+}+G^{-}
\end{equation}
where the factor of $f$ is included for convenience.

Equation ({\ref{eqn:XcontX}}) implies that the $2$-forms $J^{(i)}$ can be
regarded as anti-self-dual $2$-forms
on the base space;
\begin{equation}
\star_4 J^{(i)} = - J^{(i)}\,,
\end{equation}
where $\star_4$ denotes the Hodge dual on $B$. Moreover, they also satisfy
\begin{equation}  \label{eqn:quat}
J^{(i)}{}_m{}^p J^{(j)}{}_p{}^n = - \delta^{ij} \delta_m{}^n +
\epsilon_{ijk} J^{(k)}{}_m{}^n
\end{equation}
where indices $m,n, \ldots$ have been raised with $h^{mn}$, the inverse of
$h_{mn}$. This equation shows that the $J^{(i)}$'s satisfy the algebra of
imaginary unit quaternions, i.e., $B$ admits an almost hyper-K\"ahler
structure, just as in~\cite{gauntlett:02, gauntlett:03}.

To proceed, we use~({\ref{eqn:df}}) and~({\ref{eqn:dV}}) to obtain
\begin{eqnarray}  \label{eqn:Fsol}
X_I F^I &=&de^0 -{\frac{2 }{3}} G^+ - 2 \chi f^{-1} V_I X^I J^{(1)}  \notag
\\
&=& -f^{-1} e^0 \wedge df +{\frac{1 }{3}}G^+ + G^- -2 \chi f^{-1} V_I X^I
J^{(1)}\,.
\end{eqnarray}
From~({\ref{eqn:dPhi}}) we find that
\begin{eqnarray}  \label{eqn:nothk}
\nabla_m J^{(1)}_{np} &=& 0  \notag \\
\nabla_m J^{(2)}_{np} &=& P_m J^{(3)}_{np}  \notag \\
\nabla_m J^{(3)}_{np} &=& -P_m J^{(2)}_{np}\,,
\end{eqnarray}
where $\nabla$ is the Levi-Civita connection on $B$ and we have defined
\begin{equation}  \label{defbig}
P_m=3 \chi V_I (A^I{}_m- f X^I \omega_m )\,.
\end{equation}
From~({\ref{eqn:quat}}) and~({\ref{eqn:nothk}}) we conclude that, in the
gauged theory, the base space is K\"ahler, with K\"ahler form $J^{(1)}$. In
the ungauged theory, it is hyper-K\"ahler with K\"ahler forms $J^{(i)}$.
Again, this is all precisely as in the minimal theories~\cite{gauntlett:02,
gauntlett:03}.

We are primarily interested in the gauged theory, so we shall assume $\chi
\ne 0$. We remark however that the equations constraining the timelike class
solutions of the ungauged theory are in fact unchanged from those obtained
in \cite{gutowskireall:04} (though the constants $C_{IJK}$ were additionally
constrained in \cite{gutowskireall:04} as the scalars in that paper were
assumed to lie on a symmetric space).

Proceeding as in~\cite{gauntlett:03}, note that we can invert
({\ref{eqn:nothk}}) to solve for $P$:
\begin{equation}  \label{eqn:invrta}
P_m ={\frac{1 }{8 }} \left( J^{(3) np} \nabla_m J^{(2)}_{np}- J^{(2) np}
\nabla_m J^{(3)}_{np} \right),
\end{equation}
from which it follows that
\begin{equation}  \label{defpee}
dP=\Re\,,
\end{equation}
where $\Re$ is the Ricci-form of the base space $B$ defined by
\begin{equation}
\Re_{mn}={\frac{1 }{2}} J^{(1) pq} R_{pq mn}
\end{equation}
and $R_{pqmn}$ denotes the Riemann curvature tensor of $B$. Hence, once $B$
has been determined, $P_m$ is determined up to a gradient. An argument in
\cite{gauntlett:03} shows that the existence of $J^{(2,3)}$ obeying
equations~({\ref{eqn:quat}}) and~({\ref{eqn:nothk}}) is a consequence of $B$
being K\"ahler, and contains no further information.

Next we examine~({\ref{eqn:mostuseful}}). It is convenient to write
\begin{equation}  \label{eqn:totalident}
F^I = -f^{-1} e^0 \wedge d(fX^I) + \Psi^I + \Theta^I +X^I G^+
\end{equation}
where $\Psi^I$ is an anti-self-dual 2-form on $B$ and $\Theta^I$ is a
self-dual 2-form on $B$. Equation~({\ref{eqn:Fsol}}) implies
\begin{equation}  \label{eqn:gpluscontr}
X_I \Theta^I = -{\frac{2 }{3}} G^+
\end{equation}
and
\begin{equation}
X_I \Psi^I = G^- - 2 \chi f^{-1} V_I X^I J^{(1)}\,.
\end{equation}
Now ~({\ref{eqn:mostuseful}}) determines $\Psi^I$:
\begin{equation}  \label{eqn:psiexp}
\Psi^I = X^I G^- +{\frac{3 }{2}} \chi f^{-1} (Q^{IJ} -2 X^I X^J) V_J J^{(1)}
\end{equation}
hence
\begin{equation}  \label{eqn:rewritegaug}
F^I = d (X^I e^0) + \Theta^I +{\frac{3 }{2}} \chi f^{-1} (Q^{IJ} -2 X^I X^J)
V_J J^{(1)}\,.
\end{equation}
$\Theta^I$ is not constrained by the dilatino equation. Finally, from
({\ref{defpee}}) together with ({\ref{defbig}}) we
have the following identity
\begin{eqnarray}
3 \chi V_I \Theta^I +{\frac{9 }{2}} \chi^2 f^{-1} (Q^{IJ} -2 X^I
X^J ) V_I V_J J^{(1)} = \Re\,.
\end{eqnarray}
Contracting this expression with $J^{(1)}$ we obtain
\begin{equation}  \label{eqn:smallfsol}
f = {\frac{18 \chi^2 (Q^{IJ} -2 X^I X^J ) V_I V_J }{R}} \,,
\end{equation}
where $R$ is the Ricci scalar of $B$, and hence
\begin{equation}  \label{eqn:ricform}
\Re -{\frac{1 }{4}} R J^{(1)} = 3 \chi V_I \Theta^I\,.
\end{equation}
Finally, note that equations ({\ref{eqn:Phiproja}}) and (\ref{eqn:Phiprojb})
imply that the spinor obeys the projections
\begin{eqnarray}  \label{eqn:Jproja}
\gamma^0 \epsilon^a &=& \epsilon^a\,
\end{eqnarray}
and
\begin{eqnarray} \label{eqn:Jprojb}
J^{(1)}_{AB} \Gamma^{AB} \epsilon^a &=& - 4 \epsilon^{ab} \epsilon^b\,,
\end{eqnarray}
where indices $A,B$ refer to a vierbein $e^A$ on the base space, and
$\Gamma^A$ are gamma matrices on the base space given by $\Gamma^A = \pm i
\gamma^A$. These projections are not independent: (\ref{eqn:Jprojb}) implies
(\ref{eqn:Jproja})

So far we have been discussing constraints on the spacetime geometry and
matter fields that are necessary for the existence of a Killing spinor. We
shall now argue that these constraints are also sufficient. Assume that we
are given a metric of the form~(\ref{eqn:metric}) for which the base space
$B$ is K\"ahler. Let $J^{(1)}$ denote the K\"ahler form. Assume that $f$ is
given in terms of $X^I$ by equation~(\ref{eqn:smallfsol}) and that the field
strengths are given by equation~(\ref{eqn:rewritegaug}) where $\Theta^I$
obeys equations~(\ref{eqn:gpluscontr}) and~(\ref{eqn:ricform}). Now consider
a spinor $\epsilon^a$ satisfying the projection~(\ref{eqn:Jprojb}). It is
straightforward to show this will automatically satisfy the dilatino
equation~({\ref{eqn:newdil}}). In the basis $(e^0,f^{-1/2} e^A)$, the
gravitino equation~({\ref{eqn:grav}}) reduces to
\begin{equation}
\partial_t \epsilon^a = 0
\end{equation}
and
\begin{equation}  \label{eqn:kahlerspin}
\nabla_m \eta^a +{\frac{1 }{2}}P_m \epsilon^{ab} \eta^b=0\,,
\end{equation}
where
\begin{equation}
\eta^a=f^{-{\frac{1 }{2}}} \epsilon^a\,.
\end{equation}
The K\"ahler nature of $B$ guarantees the existence of a solution to
equation~({\ref{eqn:kahlerspin}}) obeying~(\ref{eqn:Jprojb}) without any
further algebraic restrictions~\cite{pope:82}. Therefore the above
conditions on the bosonic fields guarantee the existence of a Killing
spinor, i.e., they are both necessary and sufficient for supersymmetry. The
only projection required is~(\ref{eqn:Jprojb}), which reduces the number of
independent components of the spinor from $8$ to $2$ so we have at least
$1/4 $ supersymmetry.\footnote{%
For timelike solutions of the ungauged theory, the only projection that must
be imposed on a Killing spinor is~(\ref{eqn:Jproja}) so the solutions will
be $1/2$ supersymmetric, as in the minimal theory~\cite{gauntlett:02}.}

{\sloppy We have presented necessary and sufficient conditions for existence
of a Killing spinor. However, we are interested in supersymmetric
solutions which also satisfy the Bianchi identity $dF^I=0$ and
Maxwell equations~({\ref{eqn:gauge}}). Substituting the field strengths
(\ref{eqn:rewritegaug}) into the Bianchi identities $dF^I=0$ gives
\begin{equation}  \label{eqn:bianch}
d \Theta^I= -{\frac{3 }{2}} \chi d \big( f^{-1} (Q^{IJ} V_J -2 X^I V_J X^J)
\big) \wedge J^{(1)}\,.
\end{equation}
Note that
\begin{equation}
\star F^I = -f^{-2} \star_4 d \left(fX^I \right)+e^0 \wedge \left(
\Theta^I+X^I G^+ - \Psi^I \right),
\end{equation}
so the Maxwell equations~({\ref{eqn:gauge}}) reduce to
\begin{eqnarray}  \label{eqn:timegauge}
d \star_4 d \left( f^{-1} X_I \right) &=& -{\frac{1 }{6}}C_{IJK} \Theta^J
\wedge \Theta^K +2 \chi f^{-1} G^- \wedge J^{(1)}V_I  \notag \\
&& + {\frac{3 }{4}} \chi^2 f^{-2} \left( C_{IJK} Q^{JM} Q^{KN} V_M V_N +8
V_I V_M X^M \right) \eta_4
\end{eqnarray}
where $\eta_4$ denotes the volume form of $B$. }

Finally, the integrability conditions for the existence of a Killing spinor
guarantee that the Einstein equation and scalar equations of motion are
satisfied as a consequence of the above equations.

In summary, the general timelike supersymmetric solution is determined as
follows. First pick a K\"ahler 4-manifold $B$. Let $J^{(1)}$ denote the
K\"ahler form and $h_{mn}$ the metric on $B$. Equation~(\ref{eqn:smallfsol})
determines $f$ in terms of $X^I$. Next one has to determine $\omega$, $X^I$
and $\Theta^I$ by solving equations~(\ref{eqn:gpluscontr}),~(\ref
{eqn:ricform}),~(\ref{eqn:bianch}) and~(\ref{eqn:timegauge}) on $B$. The
metric is then given by~(\ref{eqn:metric}) and the gauge fields by~
(\ref{eqn:rewritegaug}).

\sect{The null case}

\subsection{The general solution}

In this section we shall find all solutions of the gauged $N=2$, $D=5$
supergravity for which the function $f$ vanishes everywhere.

{}From ({\ref{eqn:dV}}) it can be seen that $V$ satisfies
\be
V\wedge dV=0
\ee
 and is therefore hypersurface-orthogonal. Hence there exist functions
$u$ and $H$ such that
\begin{eqnarray}  \label{eqn:udef}
V = H^{-1} du \ .
\end{eqnarray}
In addition we find that
\begin{eqnarray}
V \cdot D V = 0 \ ,
\end{eqnarray}
so $V$ is tangent to affinely parameterized geodesics in the surfaces of
constant $u$. One can choose coordinates $(u,v,y^m)$, $m=1,2,3$, such that
$v$ is the affine parameter along these geodesics, and hence
\begin{eqnarray}
V = \frac{\partial}{\partial v}\ .
\end{eqnarray}
The metric must take the form
\begin{eqnarray}
ds^2= H^{-1} \left( \mathcal{{}F} du^2 + 2 du dv \right)-H^2 \gamma_{mn}
dy^m dy^n \ ,
\end{eqnarray}
where the quantities $H$, $\mathcal{F}$, and $\gamma_{mn}$ depend on $u$ and
$y^m$ only (because $V$ is Killing). It is particularly useful to introduce
a null basis
\begin{eqnarray}  \label{milo}
e^+ = V = H^{-1} du, \qquad e^- = dv + \frac{1}{2} \mathcal{{}F} du, \qquad
e^i= H {\hat{e}}^i
\end{eqnarray}
satisfying
\begin{eqnarray}  \label{nflip}
ds^2=2e^+e^- -e^ie^i
\end{eqnarray}
where ${\hat{e}}^i= {\hat{e}}^i_m dy^m$ is an orthonormal basis for the
3-manifold with $u$-dependent metric $\gamma_{mn}$; $\delta_{ij} {\hat{e}}^i
{\hat{e}}^j = \gamma_{mn} dy^m dy^n$.

Equation ({\ref{eqn:XcontX}}) implies that $J^{(i)}$ can
be written
\begin{eqnarray}
J^{(i)} = e^+ \wedge L^{(i)}
\end{eqnarray}
where $L^{(i)} = L^{(i)}{}_m e^m$ satisfy $L^{(i)}{}_{m} L^{(j)}{}_{n}
\delta^{mn} = \delta^{ij}$. In fact, by making a change of basis we can set
$L^{(i)}=e^i$, so
\begin{eqnarray}  \label{sgnx}
J^{(i)} = e^+ \wedge e^i = du \wedge {\hat{e}}^i \ .
\end{eqnarray}
For consistency with the computation already done for the minimal theory, we
set $\epsilon_{+-123}= -1$. Then ({\ref{eqn:cclos}}) implies
\begin{eqnarray}  \label{eqn:diferent}
du \wedge d {\hat{e}}^1 &=& 0  \notag \\
du \wedge \big[ d {\hat{e}}^2 - 3 \chi V_I (A^I \wedge {\hat{e}}^3- X^I H
{\hat{e}}^1 \wedge {\hat{e}}^2) \big] &=& 0  \notag \\
du \wedge \big[ d {\hat{e}}^3 + 3 \chi V_I(A^I \wedge {\hat{e}}^2+ X^I H
{\hat{e}}^1 \wedge {\hat{e}}^3) \big] &=& 0 \ .
\end{eqnarray}
Now define ${\tilde{d}} {\hat{e}}^i = {\frac{1 }{2}} ({\frac{\partial {\hat{e
}}^i_m }{\partial y^n}} - {\frac{\partial {\hat{e}}^i_n }{\partial y^m}})
dy^n \wedge dy^m$. Then ({\ref{eqn:diferent}}) implies that
\begin{eqnarray}  \label{eqn:diferentb}
{\tilde{d}} {\hat{e}}^1&=& 0  \notag \\
{\tilde{d}} {\hat{e}}^2 - 3 \chi V_I (A^I \wedge {\hat{e}}^3- X^I H {\hat{e}}
^1 \wedge {\hat{e}}^2) &=& 0  \notag \\
{\tilde{d}} {\hat{e}}^3 + 3 \chi V_I (A^I \wedge {\hat{e}}^2+ X^I H {\hat{e}}
^1 \wedge {\hat{e}}^3) &=& 0 \ .
\end{eqnarray}
Hence, in particular $({\hat{e}}^2+i {\hat{e}}^3) \wedge {\tilde{d}}
({\hat{e}}^2+i {\hat{e}}^3)=0$ from which it
follows that there exists a complex
function $S(u,y)$ and real functions $x^2=x^2(u,y)$, $x^3=x^3(u,y)$ such
that
\begin{eqnarray}
({\hat{e}}^2+i {\hat{e}}^3)_m = S {\frac{\partial }{\partial y^m}} (x^2+ix^3)
\end{eqnarray}
and hence $({\hat{e}}^2+i {\hat{e}}^3) = S d(x^2+ix^3)+ \psi du$ for some
complex function $\psi(u,y)$. Similarly, there exists a real function
$x^1=x^1(u,y)$ such that ${\hat{e}}^1 = dx^1 + a_1 du$ for some real function
$a_1$. Hence, from this it is clear that we can change coordinates from
$u,y^m$ to $u,x^m$. Moreover, we can make a gauge transformation of the form
$A^I \rightarrow A^I + d \Lambda^I$ where $\Lambda^I=\Lambda^I(u,x)$ in order
to set $J^{(2)}+i J^{(3)} \rightarrow S du \wedge (dx^2+idx^3)$ where $S$ is
now a real function. Note that such a gauge transformation preserves the
original gauge restriction ({\ref{eqn:gaugechoice}}) that $A^I{}_v=0$.

Hence, the null basis can be simplified to
\begin{eqnarray}  \label{eqn:milob}
e^+ = V &=& H^{-1} du  \notag \\
e^- &=& dv + \frac{1}{2} \mathcal{{}F} du  \notag \\
e^1 &=& H(dx^1+ a_1 du)  \notag \\
e^2 &=& H (Sdx^2 + S^{-1}a_2 du)  \notag \\
e^3 &=& H (S dx^3 + S^{-1} a_3 du)
\end{eqnarray}
for real functions $H(u,x^m)$, $S(u,x^m)$, $a_i (u,x^m)$, and $J^{(i)} = e^+
\wedge e^i$. We remark that this metric is that of a plane wave,
i.e. the supersymmetric Killing vector field $V$ is geodesic and free
of expansion, rotation and shear. In order for the geometry to
describe a plane-parallel wave we would also require $V$ to be
covariantly constant, which corresponds to $H$ being constant.

To proceed, we observe that equation ({\ref{eqn:df}}) implies that $i_V X_I
F^I=0$. Moreover, from ({\ref{eqn:mostuseful}}) we also find that
\begin{eqnarray}
({\frac{1 }{4}} Q_{IJ} -{\frac{3 }{8}} X_I X_J) F^J{}_{\mu \nu} (J^i)^{\mu
\nu} =0
\end{eqnarray}
for $i=1,2,3$, which can be rewritten as
\begin{eqnarray}
({\frac{1 }{4}} Q_{IJ} -{\frac{3 }{8}} X_I X_J) F^J{}_{-i}=0 \ .
\end{eqnarray}
Hence, as $X_I F^I{}_{-i}=0$ we find that $F^I{}_{-i}=0$ for $i=1,2,3$.
Next, taking ({\ref{eqn:mostusefuld}}) with $\sigma=-$, $\lambda=i$ we
obtain
\begin{eqnarray}
({\frac{1 }{4}} Q_{IJ} -{\frac{3 }{8}} X_I X_J) F^J{}_{-+}=0
\end{eqnarray}
and so, recalling that $X_I F^I{}_{-+}=0$, it follows that $F^I{}_{-+}=0$
also. Hence
\begin{eqnarray}
{}F^I = {}F^I{}_{+i} e^+ \wedge e^i + \frac{1}{2} {}F^I{}_{ij} e^i \wedge
e^j \ .
\end{eqnarray}
Now, from ({\ref{eqn:dV}}) we find
\begin{eqnarray}  \label{eqn:ffixa}
X_I F^I{}_{12} &=& H^{-2} S^{-1} \partial_3 H  \notag \\
X_I F^I{}_{13} &=& -H^{-2} S^{-1} \partial_2 H  \notag \\
X_I F^I{}_{23} &=& H^{-2} \partial_1 H - 2 \chi V_I X^I
\end{eqnarray}
where $\partial_i \equiv {\frac{\partial }{\partial x^i}}$. Next, returning
to ({\ref{eqn:mostusefulc}}) we obtain
\begin{eqnarray}  \label{eqn:ffixb}
({\frac{1 }{4}} Q_{IJ} -{\frac{3 }{8}} X_I X_J) F^J{}_{jk} \epsilon_{ijk} =
-{\frac{3 }{4}} \nabla_i X_I +{\frac{3 \chi }{2}}(X_I V_J X^J -V_I)
\delta_{i1}
\end{eqnarray}
where $\epsilon_{123}=1$. Hence, it follows from this expression together
with ({\ref{eqn:ffixa}}) that
\begin{eqnarray}  \label{eqn:FIsola}
F^I{}_{12} &=& H^{-2} S^{-1} \partial_3 H X^I +H^{-1} S^{-1} \partial_3 X^I
\notag \\
F^I{}_{13} &=& -H^{-2} S^{-1} \partial_2 H X^I -H^{-1} S^{-1} \partial_2 X^I
\notag \\
F^I{}_{23} &=& H^{-2} \partial_1 H X^I + H^{-1} \partial_1 X^I -3 \chi
Q^{IJ} V_J \ .
\end{eqnarray}
On substituting these expressions back into ({\ref{eqn:dPhi}}) we find the
following additional constraints
\begin{eqnarray}  \label{eqn:hdef}
\partial_1 S = -3 \chi V_I X^I SH
\end{eqnarray}
and
\begin{eqnarray}  \label{eqn:vfcontr}
V_I A^I = V_I A^I{}_u du +{\frac{1 }{3 \chi}} S^{-1} (\partial_2 S dx^3 -
\partial_3 S dx^2)
\end{eqnarray}
and
\begin{eqnarray}  \label{eqn:xfcontr}
X_I F^I &=& \big(-2 \chi H V_I A^I{}_u +{\frac{1 }{3}} S^{-2}
H^{-2}(\partial_2(H^3 a_3) - \partial_3 (H^3 a_2)) \big) du \wedge dx^1
\notag \\
&-&{\frac{1 }{3}} H^{-2} (\partial_1(H^3 a_3) - \partial_3 (H^3 a_1)) du
\wedge dx^2  \notag \\
&+&{\frac{1 }{3}} H^{-2} (\partial_1(H^3 a_2) - \partial_2 (H^3 a_1)) du
\wedge dx^3  \notag \\
&+& (\partial_1 H -2 \chi V_I X^I H^2) S^2 dx^2 \wedge dx^3 - \partial_2 H
dx^1 \wedge dx^3 + \partial_3 H dx^1 \wedge dx^2 \ .  \notag \\
\end{eqnarray}

It is particularly convenient to define the ``base space'' in the null
solutions to be the 3-manifold equipped with metric
\begin{eqnarray}  \label{eqn:basenull}
ds_3^2 = (dx^1)^2 + S^2 ((dx^2)^2+(dx^3)^2)
\end{eqnarray}
with positive orientation fixed by the volume form $dvol_3 = S^2 dx^1 \wedge
dx^2 \wedge dx^3$; and we denote the Hodge dual on the base space by
$\star_3 $. We denote the base metric ({\ref{eqn:basenull}}) by $h_{ij}$ with
inverse $h^{ij}$, and ${\hat{d}}$ is the exterior derivative restricted to
the base space.

Then $F^I$ can be written as
\begin{eqnarray}  \label{eqn:FImainsol}
F^I &=& du \wedge Y^I + \star_3 \big[ {\hat{d}} (H X^I) -3 \chi H^2 Q^{IJ}
V_J dx^1 \big]
\end{eqnarray}
where
\begin{eqnarray}
Y^I \equiv Y^I{}_1 dx^1 + Y^I{}_2 dx^2 + Y^I{}_3 dx^3 \ .
\end{eqnarray}
The constraint ({\ref{eqn:xfcontr}}) is equivalent to
\begin{eqnarray}  \label{eqn:xfcontrb}
X_I Y^I = {\frac{1 }{3}} H^{-2} \star_3 {\hat{d}} (H^3 a)-2 \chi H V_J
A^J{}_u dx^1
\end{eqnarray}
where $a \equiv a_1 dx^1 + a_2 dx^2 + a_3 dx^3$.

We note that the Bianchi identity implies that
\begin{eqnarray}  \label{eqn:bianchii}
{\hat{d}} \star_3 {\hat{d}} (H X^I) = 3 \chi S^{-2} \partial_1 ( S^2 H^2
Q^{IJ} ) V_J dvol_3
\end{eqnarray}
and also
\begin{eqnarray}  \label{eqn:bianchiii}
\hat{d} Y^I = \partial_u \big[ \star_3 \big({\hat{d}}(H X^I) -3 \chi H^2
Q^{IJ} V_J dx^1 \big) \big] \ .
\end{eqnarray}

In addition, the consistency conditions obtained from ({\ref{eqn:vfcontr}})
on imposing
\begin{eqnarray}
d (V_I A^I) = V_I F^I
\end{eqnarray}
are
\begin{eqnarray}  \label{eqn{eqn:vifIa}}
V_I Y^I = -{\hat{d}} (V_I A^I{}_u) +{\frac{1 }{3 \chi}} \partial_u \big(
\partial_2 \log S dx^3 - \partial_3 \log S dx^2 \big)
\end{eqnarray}
and
\begin{eqnarray}  \label{eqn:sfix}
\Box \log S = -9 \chi^2 H^2 (Q^{IJ} -2 X^I X^J) V_I V_J
\end{eqnarray}
where $\Box$ denotes the Laplacian on the base 3-manifold with metric given
by ({\ref{eqn:basenull}}).

It is also necessary to
impose the gauge equation ({\ref{eqn:gauge}}) from which we
find
\begin{eqnarray}  \label{eqn:nullgaugeeq}
{\hat{d}} (H Q_{IJ} \star_3 Y^J) &=& 3 {\hat{d}} (H^3 a) \wedge ({\frac{1 }{2
}} {\hat{d}} (H^{-1} X_I)+ \chi V_I dx^1)  \notag \\
&+& {\frac{1 }{2}} C_{IJK} \star_3 Y^J \wedge ({\hat{d}} (H X^K) -3 \chi H^2
Q^{KP} V_P dx^1) \ .
\end{eqnarray}

Next, we recall that imposing supersymmetry together with the gauge
equations implies that all of the components of the Einstein equation are
satisfied automatically with the exception of the $++$ component, which must
be evaluated independently. In particular,
\begin{eqnarray}
R_{++} = {\frac{1 }{2H}} \Box {\mathcal{F}} + \partial_u {\hat{W}} - a^i
\partial_i {\hat{W}} - W_{ij} W^{ij}
\end{eqnarray}
where
\begin{eqnarray}
W_{ij} = H {\hat{\nabla}}_{(i} a_{j)} + (a^k \partial_k H - \partial_u H)
h_{ij} - {\frac{1 }{2}} H \partial_u h_{ij}
\end{eqnarray}
and ${\hat{W}}= W^i{}_i$, indices on $W$ and $a$ are raised with $h^{ij}$,
the inverse of the base metric and ${\hat{\nabla}}$ denotes the Levi-Civita
connection on the base manifold. It is convenient to define 1-forms $U^I$ on
the base space by
\begin{eqnarray}
U^I \equiv Y^I + \star_3 \big[ a \wedge \big({\hat{d}}(H X^I) -3 \chi H^2
Q^{IJ} V_J dx^1 \big) \big]
\end{eqnarray}
and for any function $g(u,x^i)$ we define
\begin{eqnarray}
{\mathcal{D}} g \equiv \partial_u g - a^i \partial_i g \ .
\end{eqnarray}
Then ${\mathcal{F}}$ is constrained via
\begin{eqnarray}
\Box {\mathcal{F}} = -2H ({\mathcal{D}} {\hat{W}} -W_{ij} W^{ij}) +2H Q_{IJ}
\big( U^I_i U^{Ji} +H^2 {\mathcal{D}} X^I {\mathcal{D}} X^J \big)
\end{eqnarray}
where the indices of $U^I$ have been raised with $h^{ij}$.

{}Finally, it remains to substitute the bosonic constraints into the
gravitino equation ({\ref{eqn:grav}}) and to check that the geometry does
indeed admit Killing spinors. If we impose the constraint
\begin{eqnarray}  \label{eqn:annihksp}
\gamma^+ \epsilon^a =0
\end{eqnarray}
on the Killing spinor, then the $\alpha=-$ component of the Killing spinor
equation implies that
\begin{eqnarray}  \label{eqn:liedera}
{\frac{\partial \epsilon^a }{\partial v}}=0 \ ,
\end{eqnarray}
so $\epsilon^a = \epsilon^a (u,x^1,x^2,x^3)$. Next we set $\alpha=+$; we
find that
\begin{eqnarray}  \label{eqn:dplus}
H \big( \partial_u \epsilon^a - a^i \partial_i \epsilon^a \big) -{\frac{\chi
}{2}} V_I X^I \gamma^- (\gamma^1 \epsilon^a + \epsilon^{ab} \epsilon^b) +{
\frac{3 \chi }{2}} V_I A^I{}_+ (\gamma^1 \epsilon^a + \epsilon^{ab}
\epsilon^b) =0 \ ,
\end{eqnarray}
where we have raised the indices on $a_i$ using the base metric ({\ref
{eqn:basenull}}). Acting on ({\ref{eqn:dplus}}) with $\gamma^+$ we find the
algebraic constraint
\begin{eqnarray}  \label{eqn:extraconstr}
\gamma^1 \epsilon^a + \epsilon^{ab} \epsilon^b=0 \ .
\end{eqnarray}
Next set $\alpha=1,2,3$; it is straightforward to show that these components
of the Killing spinor equation imply that
\begin{eqnarray}  \label{eqn:constabc}
\partial_1 \epsilon^a = \partial_2 \epsilon^a = \partial_3 \epsilon^a =0
\end{eqnarray}
and substituting this back into ({\ref{eqn:dplus}}) we also find
\begin{eqnarray}  \label{eqn:dpplus}
\partial_u \epsilon^a =0 \ .
\end{eqnarray}
Hence the gravitino equation implies that $\epsilon^a$ is constant.
Moreover, it is straightforward to check that the dilatino equation
({\ref{eqn:dil}}) is satisfied.

It is also useful to examine the effect on the solution of certain
co-ordinate transformations. In particular, under the shift $v=v^{\prime}+
g(u,x)$ we note that the form of the solution remains the same, with $v$
replaced by $v^{\prime}$, and $a_i$ and ${\mathcal{F}}$ replaced by
\begin{eqnarray}  \label{eqn:coordsimp}
a^{\prime}{}_i &=& a_i - H^{-3} \nabla_i g  \notag \\
{\mathcal{F}}^{\prime}&=& {\mathcal{F}} +2 {\frac{\partial g }{\partial u}}
-2 a^i \partial_i g +H^{-3} h^{ij} \partial_i g \partial_j g \ .
\end{eqnarray}
Hence we see that $H^3 a$ is determined only up to a gradient.

\subsection{Solutions with $F^I=0$}

It is particularly instructive to examine the solutions for which $F^I=0$.
To begin, observe that the vanishing of $F^I{}_{12}$ and $F^I{}_{13}$ in
({\ref{eqn:FIsola}}) implies that
\begin{eqnarray}
\partial_2 (HX^I) = \partial_3 (HX^I)=0
\end{eqnarray}
and hence
\begin{eqnarray}
\partial_2 H = \partial_3 H = \partial_2 X^I = \partial_3 X^I =0
\end{eqnarray}
so $H$ and $X^I$ depend only on $x^1$ and $u$. Then, from equation
({\ref{eqn:hdef}}) it follows that $S$ must be separable as
\begin{eqnarray}
S = S_1 (u,x^1) S_2 (u,x^2,x^3) \ .
\end{eqnarray}
Note that $V_I Y^I=0$ implies that $\partial_u \partial_2 \log S_2 =
\partial_u \partial_3 \log S_2 =0$. Hence without loss of generality, we can
set $S_2 = S_2 (x^2,x^3)$. Moreover, from the vanishing of $V_I F^I$, using
({\ref{eqn:vfcontr}}) we find that
\begin{eqnarray}
(\partial_2^2 + \partial_3^2) \log S_2 =0 \ ,
\end{eqnarray}
hence by making an appropriate ($u$ and $x^1$-independent) change of $x^2,
x^3$ co-ordinates together with an $u$, $x^1$-independent gauge
transformation, we can without loss of generality take $S_2=1$ and set
$S=S(u,x^1)$. Note that in these new co-ordinates ({\ref{eqn:vfcontr}}) also
implies that
\begin{eqnarray}
V_I A^I{}_u = P(u)
\end{eqnarray}
for some function $P(u)$.

Next, from the vanishing of $F^I{}_{23}$ in ({\ref{eqn:FIsola}}), we find
that
\begin{eqnarray}  \label{eqn:xupeqn}
\partial_1 (H X^I) = 3 \chi Q^{IJ} V_J H^2 \ .
\end{eqnarray}
Contracting this equation with $X_I$ we obtain
\begin{eqnarray}  \label{eqn:hsfix}
\partial_1 H = 2 \chi X^I V_I H^2 \ .
\end{eqnarray}
Then ({\ref{eqn:hdef}}) together with ({\ref{eqn:hsfix}}) imply that
\begin{eqnarray}
H = S^{-{\frac{2 }{3}}} Q(u)
\end{eqnarray}
for some function $Q(u)$. However, by making a redefinition of the
co-ordinate $u$ we can without loss of generality take $Q=1$ and so
\begin{eqnarray}
H = S^{-{\frac{2 }{3}}} \ .
\end{eqnarray}
Also, note that ({\ref{eqn:xupeqn}}) is equivalent to
\begin{eqnarray}
\partial_1 (H^{-1} X_I) = -2 \chi V_I
\end{eqnarray}
which we solve by setting
\begin{eqnarray}
H^{-1} X_I = -2 \chi V_I x^1 + \beta_I (u)
\end{eqnarray}
for some functions $\beta_I (u)$. If the scalars lie on a symmetric space,
then this equation can be inverted to obtain an explicit solution for $H$.

Next consider the $a_i$. By making a co-ordinate transformation of the type
given in ({\ref{eqn:coordsimp}}) we can set $a_1=0$. Then from the
constraint ({\ref{eqn:xfcontrb}}) we find that $H^3 a_2$ and $H^3 a_3$ are
independent of $x^1$. In addition, setting $b_2 = H^3 a_2$, $b_3 = H^3 a_3$
we must satisfy
\begin{eqnarray}
\partial_2 b_3 - \partial_3 b_2 = 6 \chi P \ .
\end{eqnarray}
This fixes $b_2$, $b_3$ up to an arbitrary gradient of a function of $x^2$,
$x^3$ and $u$, this gradient can also be removed by using a co-ordinate
transformation of the type given in ({\ref{eqn:coordsimp}}). We can
therefore set without loss of generality
\begin{eqnarray}
b_2 = -3 \chi P x^3 , \qquad b_3 = 3 \chi P x^2 \ .
\end{eqnarray}
It is convenient to set $x^2= r \cos \theta$, $x^3=r \sin \theta$, then the
metric can be rewritten as
\begin{eqnarray}
ds^2 = H^{-1} \big[2 du(dv +{\frac{1 }{2}} {\mathcal{F}} du)-dr^2 - r^2(d
\theta + 3 \chi P du)^2 \big] -H^2 (dx^1)^2 \ .
\end{eqnarray}
So, by making a shift in $\theta$ we can without loss of generality set $P=0$
, and the metric can be simplified to
\begin{eqnarray}
ds^2 = H^{-1} \big[2 du(dv +{\frac{1 }{2}} {\mathcal{F}} du)-(dx^2)^2 -
(dx^3)^2 \big] -H^2 (dx^1)^2 \ .
\end{eqnarray}

Lastly, it remains to impose the Einstein equations which fix ${\mathcal{F}}$
via
\begin{eqnarray}
H \partial_1^2 {\mathcal{F}} + H^4 (\partial_2^2 + \partial_3^2) {\mathcal{F}
} -3 \partial_1 H \partial_1 {\mathcal{F}} = {\frac{9 }{2}} H^6 Q^{IJ}
\partial_u \beta_I \partial_u \beta_J \ .
\end{eqnarray}
Hence we see that solutions for which ${\mathcal{F}}=0$ must have $%
\partial_u \beta_I=0$ and hence $H$ and $X^I$ are also independent of $u$.

\subsection{Magnetic Null Solutions}

In order to find some more general solutions with $F^I \neq 0$ we set
\begin{eqnarray}
F^I = B^I dx^2 \wedge dx^3
\end{eqnarray}
for some functions $B^I$, we shall assume that the $B^I$ do not all vanish.
From the Bianchi identity we must have $B^I = B^I (x^2 , x^3)$. Then it is
straightforward to see that for exactly the same reasoning as for the
solutions with $F^I=0$, one must have $H=H(u,x^1)$ and $X^I = X^I (u,x^1)$
with $S$ separable as
\begin{eqnarray}
S = S_1(u,x^1) S_2(x^2,x^3)
\end{eqnarray}
and the vanishing of $V_I Y^I$ implies
\begin{eqnarray}
V_I A^I{}_u = P(u)
\end{eqnarray}
for some function $P$. The vanishing of $Y^I$ also implies from ({\ref
{eqn:xfcontrb}}) that
\begin{eqnarray}
{\frac{1 }{3}} H^{-2} \star_3 {\hat{d}} (H^3a) -2 \chi HP dx^1 =0 \ .
\end{eqnarray}

Next, observe that ({\ref{eqn:FImainsol}}) implies that
\begin{eqnarray}  \label{eqn:Bisol}
B^I = S^2 \big( \partial_1 (HX^I)-3 \chi H^2 Q^{IJ} V_J \big) \ .
\end{eqnarray}
Then, from the gauge equations ({\ref{eqn:nullgaugeeq}}) we find that
\begin{eqnarray}
Q_{IJ} B^J {\hat{d}} (H^3 a) \wedge dx^1 =0 \ .
\end{eqnarray}
As $Q_{IJ} B^J$ do not all vanish, it follows that
\begin{eqnarray}
{\hat{d}} (H^3 a)=0
\end{eqnarray}
and therefore $P=0$. By making use of a co-ordinate transformation of the
form ({\ref{eqn:coordsimp}}) we can also without loss of
generality set $a=0$.

From ({\ref{eqn:Bisol}}) it follows that
\begin{eqnarray}
(S_2)^{-2} B^I = (S_1)^2 (\partial_1 (H X^I) -3 \chi H^2 Q^{IJ} V_J) \ .
\end{eqnarray}
The LHS of this expression is a function of $x^2, x^3$ alone, whereas the
RHS depends on $u$ and $x^1$ only. Hence we must have
\begin{eqnarray}
B^I = (S_2)^2 q^I
\end{eqnarray}
for some constants $q^I$. Then the gauge field strengths simplify to
\begin{eqnarray}
F^I = q^I \mathrm{dvol} (M_2)
\end{eqnarray}
where $M_2$ is the 2-manifold equipped with metric
\begin{eqnarray}
ds^2 (M_2) = (S_2)^2 \big( (dx^2)^2 + (dx^3)^2 \big) \ .
\end{eqnarray}
Then from ({\ref{eqn:vfcontr}}), requiring that $d(V_I A^I)=V_I F^I$ we find
that $S_2$ must satisfy
\begin{eqnarray}
(\partial_2^2+\partial_3^2) \log S_2 = 3 \chi V_I q^I (S_2)^2
\end{eqnarray}
which is the Liouville equation. Hence, by making a ($u$, $x^1$ independent)
co-ordinate transformation of $x^2$, $x^3$ we can take the metric on
$M_2$ to be that of $H^2$, $\bR^2$ or $S^2$ according as
$\chi V_I q^I >0$, $\chi V_I q^I =0$ or $\chi V_I q^I <0$ respectively.

Also, $H$, $S_1$ and $X^I$ are constrained by
\begin{eqnarray}
\partial_1 S_1 = -3 \chi V_I X^I S_1 H
\end{eqnarray}
and
\begin{eqnarray}
H^2 (S_1)^2 \partial_1 \big[ H^{-1} X_I +2 \chi V_I x^1 \big] = -{\frac{2 }{3
}} Q_{IJ} q^J \ .
\end{eqnarray}
In general, it is not possible to integrate up these equations when
$q^I \neq 0$. When the scalar manifold is particularly simple, such as
in the ``STU'' model, solutions (though by no means the most general
solution) to these equations have been found \cite{cks1}. In principle,
the equations presented here could be used to find $u$-dependent
generalizations of these solutions; though we shall not
pursue this further here.

\subsection{Null Solutions of the Ungauged Theory}

Having examined the null solutions of the gauged theory, it is
straightforward to consider the special case when the gauge parameter
vanishes. There is then considerable simplification to many of the equations.

In particular, as the $J^{(i)}$ are all now closed, we can introduce
co-ordinates $x^i$ for $i=1,2,3$ such that
\begin{eqnarray}
J^{(i)}= e^+ \wedge e^i = du \wedge dx^i
\end{eqnarray}
and so we can take for a null basis
\begin{eqnarray}  \label{eqn:milobb}
e^+ = V &=& H^{-1} du  \notag \\
e^- &=& dv + \frac{1}{2} \mathcal{{}F} du  \notag \\
e^i &=& H(dx^i+ a_i du) \quad i=1,2,3 \ .
\end{eqnarray}

It is then straightforward to show that the differential constraints on the
spinor bilinears together with the algebraic constraints on the bilinears
obtained from the dilatino equation imply that
\begin{eqnarray}  \label{eqn:ungaugedFsol}
F^I = du \wedge Y^I + \star_3 {\hat{d}} (HX^I)
\end{eqnarray}
where here $\star_3$ denotes the Hodge dual on $\mathbb{R}^3$ equipped with
the standard metric
\begin{eqnarray}
ds^2 (\mathbb{R}^3)= (dx^1)^2+(dx^2)^2+(dx^3)^2
\end{eqnarray}
and positive orientation is defined with respect to the volume form $dx^1
\wedge dx^2 \wedge dx^3$; and ${\hat{d}}$ is the exterior derivative
restricted to $\mathbb{R}^3$. The $Y^I$ are 1-forms on $\mathbb{R}^3$ which
must satisfy
\begin{eqnarray}  \label{eqn:ungaugedha}
X_I Y^I = {\frac{1 }{3}} H^{-2} \star_3 {\hat{d}} (H^3a) \ .
\end{eqnarray}

The Bianchi identity implies that
\begin{eqnarray}  \label{eqn:ungaugedcurl}
{\hat{d}} Y^I = \star_3 {\hat{d}} \big( \partial_u (HX^I) \big)
\end{eqnarray}
together with
\begin{eqnarray}
\nabla^2 (H X^I)=0
\end{eqnarray}
where $\nabla^2$ is the Laplacian on $\mathbb{R}^3$. Hence we can write
\begin{eqnarray}  \label{eqn:ungaugedxidef}
X^I = H^{-1} K^I
\end{eqnarray}
where $K^I$ are $u$-dependent harmonic functions on $\mathbb{R}^3$. $H$ is
then fixed in terms of these harmonic functions by
\begin{eqnarray}  \label{eqn:ungaugedhdef}
H^3 = {\frac{1 }{6}} C_{IMN} K^I K^M K^N \ .
\end{eqnarray}

Next consider the gauge equations, which can be written using the above
constraints as
\begin{eqnarray}  \label{eqn:gaugerewrite}
C_{IMN} \big( Y^M{}_i \nabla^i K^N +{\frac{1 }{2}} \nabla^i Y^M{}_i K^N \big)
=0
\end{eqnarray}
where here $\nabla_i \equiv {\frac{\partial }{\partial x^i}}$ and indices
are raised with $\delta^{ij}$.

Finally, we note that supersymmetric solutions of the ungauged theory are
generically ${\frac{1 }{2}}$-supersymmetric, in contrast to the ${\frac{1 }{4
}}$ supersymmetric solutions of the gauged theory. It is straightforward to
show that on substituting the constraints into the Killing spinor equation
we find that
\begin{eqnarray}
\partial_\mu \epsilon^a =0
\end{eqnarray}
so that $\epsilon^a$ is constant, and is constrained by
\begin{eqnarray}
\gamma^+ \epsilon^a =0 \ .
\end{eqnarray}

It is clear that the equations for the ungauged solutions decouple much more
straightforwardly than for gauged solutions. In particular, to construct a
solution one first chooses $u$-dependent harmonic functions $K^I$, and
defines $H$ by ({\ref{eqn:ungaugedhdef}}) and then $X^I$ are given by
({\ref{eqn:ungaugedxidef}}).
Now as $\partial_u K^I$ are also harmonic functions
on $\mathbb{R}^3$, it follows locally that we can always find a 1-form $Y^I$
on $\mathbb{R}^3$ satisfying ({\ref{eqn:ungaugedcurl}}), though this
equation only fixes $Y^I$ up to a gradient.

Suppose that ${\tilde{Y}}^I$ is a particular integral of
({\ref{eqn:ungaugedcurl}}); then we can write
\begin{eqnarray}  \label{eqn:yiclos}
Y^I = d Z^I + {\tilde{Y}}^I
\end{eqnarray}
for some functions $Z^I$. Then ({\ref{eqn:gaugerewrite}}) implies
\begin{eqnarray}  \label{eqn:particularint}
\nabla^2 (C_{IMN} Z^M K^N) = - C_{IMN} \big( 2 {\tilde{Y}}^M{}_i \nabla^i
K^N + \nabla^i {\tilde{Y}}^M{}_i K^N \big)
\end{eqnarray}
which fixes $C_{IMN} Z^M K^N$ up to some other ($u$-dependent) harmonic
functions on $\mathbb{R}^3$.

Then, given such $Y^I$, on contracting ({\ref{eqn:gaugerewrite}}) with $K^I$
we obtain the condition
\begin{eqnarray}
{\hat{d}} \star_3 (H^2 X_I Y^I)=0 \ .
\end{eqnarray}
It follows that there exists $H^3 a$ satisfying ({\ref{eqn:ungaugedha}});
which is fixed up to an arbitrary gradient. This gradient can be removed by
making a shift in the co-ordinate $v$ of the type
given in ({\ref{eqn:coordsimp}}). Finally, it
remains to solve the ++ component of the
Einstein equations which fix ${\mathcal{F}}$ up to another $u$-dependent
harmonic function on $\mathbb{R}^3$. Hence, it is clear that the whole
solution is specified completely in terms of harmonic functions on $\mathbb{R%
}^3$. Moreover, it is also apparent that there is considerable
simplification when the harmonic functions $K^I$ are independent of $u$; as
in this case it follows that $H$ is also independent of $u$, and one can set
${\tilde{Y}}^I=0$ in ({\ref{eqn:yiclos}}) and ({\ref{eqn:particularint}}).
Such solutions were constructed in \cite{chsabra:1999}.

\sect{Conclusions}

In this paper we have completed the classification of supersymmetric
solutions of $N=2$, $D=5$ gauged supergravity
which preserve  2 of the 8 supersymmetries.
It is known that the only solution which preserves all 8
of the supersymmetries is $AdS_5$ with vanishing gauge field strengths
$F^I=0$ and constant scalars $X^I$. It would be interesting to determine
the structure of solutions preserving 1/2 or 3/4 of the supersymmetries.
One expects the geometries of these solutions to be constrained
further by the presence of additional supersymmetry.
It has been shown, for a class of solutions in $N=2$, $D=4$ supergravity
for which the Killing vector obtained from the Killing spinor is null,
that there are no 3/4 supersymmetric solutions \cite{klemmcald:04}.
The status of 3/4 supersymmetric solutions in the timelike class of
the theories examined in \cite{klemmcald:04}  has yet
to be determined. This situation is in contrast to that
encountered in the ungauged theories, where all supersymmetric solutions
are either half-supersymmetric or maximally supersymmetric.

Unfortunately, it appears to be rather awkward to
adapt the spinor-bilinear approach to investigate solutions
with additional supersymmetries.
A more promising method for dealing with such solutions has recently
been developed in the context of eleven dimensional supergravity
in \cite{papadop:04}, extending the previous classifications of
solutions of this theory \cite{gauntlett:03a}, \cite{gauntlett:03b}.
Although both methods are dealing with the same problem, the
approach in \cite{papadop:04} is advantageous for investigating
solutions with enhanced supersymmetry, because it allows for a particularly
explicit realization of the Killing spinor in terms of differential forms.
For, although the spinor-bilinear approach adopted here is
reasonably straightforward in low-dimensional supergravities,
it becomes extremely complicated in higher dimensions. This is because
one must make use of Fierz identities in order to compute the
algebraic relations between the bilinears, and when there is more than
one Killing spinor, there are many bilinears.

The equations which we have obtained in this paper are considerably more
complicated than those obtained in the classifications of the minimal
gauged and ungauged supergravities. This is not surprising, as
solutions of the minimal theory form a very restricted class of
solutions when lifted to higher dimensions. Including more multiplets in
the lower dimensional theory corresponds to considering more generic
solutions in higher dimensions. The constraints obtained from
just the Killing spinor equations (not considering the Bianchi
or gauge equations) for 1/32 supersymmetric eleven-dimensional solutions
in  \cite{gauntlett:03a}, \cite{gauntlett:03b} are extremely complicated.
Hence we expect solutions of the five-dimensional supergravities
with additional vector multiplets included to reflect some of this
increased complexity. However, as many eleven-dimensional
solutions which are of physical interest have more than
1/32 supersymmetry, it is to be hoped that classifications of
these solutions can be obtained which have meaningful geometric properties.

A final outstanding question is whether there are
any regular asymptotically AdS black ring solutions.
Supersymmetric black rings are known to exist in the ungauged theory
\cite{harveyelv:04, harveyelv:04b, warner:04a, ggbr1:04, ggbr2:04}.
As an asymptotically flat black hole can be obtained from
these ring solutions by tuning
a particular parameter to vanish, it is natural to enquire whether
the black hole solutions found in \cite{gutowski:04}
can be regarded as a special case of a family of AdS black rings.
It is clear that a better understanding of the near-horizon
geometries of solutions with regular horizons could be used to
exclude this possibility. Moreover, as there is supersymmetry enhancement
near the horizons of ungauged black holes and rings, an examination of
1/2-supersymmetric  gauged solutions could be valuable.

\medskip

\acknowledgments{J.B.G. thanks the Perimeter Institute for support, at which part
of this work was completed. W.S. thanks the Mathematical
Institute, Oxford, for hospitality during the early stages of this
work. The work of W.S. is supported in part by NSF grant
PHY-0313416.}

%%%%%%%%%%%%%%%%%%%%


\begin{thebibliography}{99}

\bibitem{maldacena:98} J.~M.~Maldacena, \textit{The large N limit of
superconformal field theories and supergravity, Adv.\ Theor.\ Math.\ Phys.\ }
\textbf{2} (1998) 231; [\textit{Int.\ J.\ Theor.\ Phys.\ } \textbf{38}
(1999) 1113] [hep-th/9711200].

\bibitem{witten:98} E.~Witten, \textit{Anti-de Sitter space and holography,
Adv. Theor. Math. Phys. } \textbf{2} (1998) 253 [hep-th/9802150].

\bibitem{oz:98} O. Aharony, S. S. Gubser, J. M. Maldacena, H. Ooguri and
Y. Oz, \textit{Large N field
theories, string theory and gravity, Phys.\ Rept.}
\ \textbf{323} (2000) 183 [hep-th/9905111].

\bibitem{deBoer:1999xf} {\normalsize J.~de Boer, E.~Verlinde and
H.~Verlinde, \textit{On the holographic renormalization group}, \textit{JHEP}
\textbf{0008} (2000) 003 [hep-th/9912012]. }

\bibitem{hp} S. W. Hawking and D. N. Page, \textit{Thermodynamics of black
holes in anti-de Sitter space},\textit{\ Commun. Math. Phys..} \textbf{87}
(1983) 577.

\bibitem{wit} E. Witten, \textit{Anti-de Sitter space, thermal phase
transition, and confinement in gauge theories,
Adv. Theor. Math. Phys.} \textbf{2} (1998) 505.

\bibitem{cs1} A. H. Chamseddine and W. A. Sabra, \textit{Magnetic Strings In
Five Dimensional Gauged Supergravity Theories}, \textit{\ Phys. Lett}.
\textbf{B477 }(2000) 329 [hep-th/9911195].

\bibitem{ks1} D. Klemm and W. A. Sabra, \textit{Supersymmetry of Black
Strings in D=5 Gauged Supergravities}, \textit{Phys. Rev.} \textbf{D62}
(2000) 024003 [hep-th/0001131].

\bibitem{behrndt:98} K.~Behrndt, A.~H.~Chamseddine and W.~A.~Sabra, \textit{
BPS black holes in N = 2 five dimensional AdS supergravity, Phys.\ Lett. }
\textbf{B442} (1998) 97 [hep-th/9807187].

\bibitem{klemm:01a} D.~Klemm and W.~A.~Sabra, \textit{General (anti-)de
Sitter black holes in five dimensions, JHEP} \textbf{0102} (2001) 031
[hep-th/0011016].

\bibitem{cks1} S. L. Cacciatori, D. Klemm and W. A. Sabra, \textit{
Supersymmetric Domain Walls and Strings in D=5 gauged Supergravity coupled
to Vector} \textit{Multiplets}, \textit{\ JHEP} \textbf{0303} (2003) 023
[hep-th/0302218].

\bibitem{gauntlett:03} J.~P.~Gauntlett and J.~B.~Gutowski, \textit{All
supersymmetric solutions of minimal gauged supergravity in five dimensions,
Phys.\ Rev. } \textbf{D68} (2003) 105009.

\bibitem{tod:83} K.~P.~Tod, \textit{All Metrics Admitting Supercovariantly
Constant Spinors, Phys. Lett. } \textbf{B121} (1983) 241.

\bibitem{gutowski:04} J.~B.~Gutowski and H.~S.~Reall, \textit{Supersymmetric
$AdS_{5}$ black holes, JHEP} \textbf{02} (2004) 006 [hep-th/0401042].

\bibitem{gutowskireall:04} J. B. Gutowski and H. S. Reall, \textit{General
Supersymmetric $AdS_{5}$ Black Holes,} JHEP 0404 (2004) 048 [hep-th/0401129].

\bibitem{cveticpope1:05} Z. W. Chong, M. Cvetic, H. Lu and  C. N. Pope,
\textit{Five-Dimensional Gauged Supergravity Black Holes with
Independent Rotation Parameters} [hep-th/0505112].

\bibitem{cveticpope2:05} Z. W. Chong, M. Cvetic, H. Lu and  C. N. Pope,
\textit{General Non-Extremal Rotating Black Holes in Minimal
Five-Dimensional Gauged Supergravity} [hep-th/0506029].

\bibitem{gunaydin:85} M.~Gunaydin, G.~Sierra and P.~K.~Townsend, \textit{
Gauging The D = 5 Maxwell-Einstein Supergravity Theories: More On Jordan
Algebras, Nucl. Phys. } \textbf{B253} (1985) 573.

\bibitem{gauntlett:02} J.~P.~Gauntlett, J.~B.~Gutowski, C.~M.~Hull, S.~Pakis
and H.~S.~Reall, \textit{All supersymmetric solutions of minimal
supergravity in five dimensions, Class.\ Quant.\ Grav. } \textbf{20} (2003)
4587 [hep-th/0209114].

\bibitem{reall:03} H.~S.~Reall, \textit{Higher dimensional black holes and
supersymmetry, Phys.\ Rev. } \textbf{D68} (2003) 024024 [hep-th/0211290].

\bibitem{pope:82} C.~N.~Pope, \textit{Kahler Manifolds And Quantum Gravity,
J.\ Phys. } \textbf{A15} (1982) 2455.

\bibitem{chsabra:1999}  A. H. Chamseddine and W. A. Sabra,
\textit{Calabi-Yau Black Holes and Enhancement of Supersymmetry in
Five Dimensions, Phys. Lett.} \textbf{B460} (1999) 63 [hep-th/9903046].

\bibitem{klemmcald:04} S. L. Cacciatori, M. M. Caldarelli, D. Klemm and
D. S. Mansi, \textit{More on BPS solutions of N=2, D=4 gauged supergravity,
JHEP} {\textbf{0407}} (2004) 061 [hep-th/0406238].

\bibitem{papadop:04} J. Gillard, U. Gran and G. Papadopoulos,
\textit{The spinorial geometry of supersymmetric backgrounds,
Class. Quant. Grav.} {\textbf{22}} (2005) 1033 [hep-th/0410155].

\bibitem{gauntlett:03a} J.~P.~Gauntlett and S.~Pakis, \textit{The geometry
of D = 11 Killing spinors, JHEP} \textbf{0304} (2003) 039 [hep-th/0212008].

\bibitem{gauntlett:03b} J.~P.~Gauntlett, J.~B.~Gutowski and S.~Pakis,
\textit{The geometry of D = 11 null Killing spinors, JHEP} \textbf{0312}
(2003) 049 [hep-th/0311112].

\bibitem{harveyelv:04} H. Elvang, R. Emparan, D. Mateos and H. S. Reall,
\textit{ A supersymmetric black ring, Phys. Rev. Lett.} {\textbf{93}}
(2004) 211302 [hep-th/0407065].

\bibitem{harveyelv:04b}  H. Elvang, R. Emparan, D. Mateos and H. S. Reall,
\textit{Supersymmetric black rings and three-charge supertubes,
Phys. Rev.} {\textbf{D71}} (2005) 024033 [hep-th/0408120].

\bibitem{warner:04a} I. Bena and N. P. Warner, \textit{
One Ring to Rule Them All ... and in the Darkness Bind Them?}
[hep-th/0408106].

\bibitem{ggbr1:04} J. P. Gauntlett and J. B. Gutowski, \textit{
Concentric Black Rings, Phys. Rev.} {\textbf{D71}} (2005) 025013
[hep-th/0408010].

\bibitem{ggbr2:04} J. P. Gauntlett and J. B. Gutowski, \textit{
General Concentric Black Rings, Phys. Rev.} {\textbf{D71}}
(2005) 045002 [hep-th/0408122].



\end{thebibliography}
\end{document}